\documentclass[reprint,prl,aps,superscriptaddress,amsmath,longbibliography]{revtex4-2}
\usepackage{graphicx,txfonts,dcolumn}
\usepackage[colorlinks,linkcolor=blue,citecolor=blue,anchorcolor=blue,urlcolor=blue]{hyperref}

\begin{document}

\title{Multiple and Asymmetric Scalings in Explosive Percolation}

\author{Ming Li}
\email{lim@hfut.edu.cn}
\affiliation{Department of Physics, Hefei University of Technology, Hefei, Anhui 230009, P. R. China}

\author{Junfeng Wang}
\affiliation{Department of Physics, Hefei University of Technology, Hefei, Anhui 230009, P. R. China}

\author{Youjin Deng}
\email{yjdeng@ustc.edu.cn}
\affiliation{Hefei National Laboratory for Physical Sciences at Microscale and Department of Modern Physics, University of Science and Technology of China, Hefei, Anhui 230026, P. R. China}
\affiliation{MinJiang Collaborative Center for Theoretical Physics, Department of Physics and Electronic Information Engineering, Minjiang University, Fuzhou, Fujian 350108, P. R. China}

\date{\today}

\begin{abstract}
Explosive percolation in the Achlioptas process has recently
attracted much research attention. From extensive simulations
in an event-based ensemble, we find that, in dimensions from $2$
to $6$ and on random graphs, the Achlioptas processes all
have two scaling windows and multiple fractal structures.
The mixing of these multiple scalings
successfully explains the previously observed anomalous
phenomena in the conventional ensemble,
and, moreover, correct critical exponents
are now determined with a high precision by the event-based method.
The multiple scalings and the ensemble inequivalence
may bring new insights for other statistical systems.
\end{abstract}

\maketitle

Percolation is one of the paradigms in statistical physics
and probability theory~\cite{Stauffer1991}.
The standard percolation model on a lattice is defined
by randomly occupying sites or bonds with an independent
probability, and undergoes a continuous phase transition.
A simple alteration of the percolation, such as lattice type,
only results in a different critical point,
and does not change the universality class~\cite{Stauffer1991}.
By adopting significantly different percolation rules,
such as rigidity percolation~\cite{Jacobs1995,Jacobs1996},
new universalities can arise.
Further, by introducing the dimension of time,
the directed percolation~\cite{Grassberger1989,Wang2013}
converts the percolation to a dynamical process
and the transition becomes a non-equilibrium type with
asymmetric exponents in spatial and temporal dimensions.
Nevertheless, the continuity of the transition remains robust.

In recent years, there is an ongoing discussion on
the so-called Achlioptas process~\cite{Boccaletti2016},
in which some intrinsic mechanism is introduced
to suppress the growth of large clusters
so that the percolation transition is delayed
and becomes explosively sharp. A basic way is called
the product rule~\cite{Achlioptas2009}.
At each time step, a random pair of empty bonds is picked up,
the size product of the two clusters containing the
ending sites of each bond is calculated,
and the one, leading to a smaller size product, is inserted.
As a consequence, the onset of percolation is significantly
delayed, but once it happens, large clusters emerge
suddenly, hence the name explosive percolation.
A wide class of Achlioptas processes with
explosive transition has been observed~\cite{Boccaletti2016},
including on regular lattices~\cite{Ziff2009,Ziff2010}
and scale-free networks~\cite{Cho2009,Radicchi2009}, and
systems with other percolation
rules~\cite{Friedman2009,Costa2010,DSouza2010,Nagler2011,Riordan2012}.

Along with the sudden appearance of large clusters,
the explosive percolation shows rich phenomena which
are often recognized as signs of a discontinuous phase transition,
such as the powder keg mechanism~\cite{Friedman2009},
bimodal distribution of the order parameter~\cite{Grassberger2011,Tian2012},
non-self-averaging property~\cite{Riordan2012},
and hysteresis~\cite{Bastas2011}.
Actually, the explosive percolation was perceived as
a discontinuous transition when it was
introduced~\cite{Achlioptas2009,Friedman2009,Ziff2009,Cho2009,Radicchi2009,Ziff2010,Radicchi2010,DSouza2010,Cho2011}.
However, later studies suggested that the sharp transition
is continuous, but exhibits anomalous critical
behaviors~\cite{Costa2010,Lee2011,Grassberger2011,Riordan2011,DSouza2015}.
A theoretical study~\cite{Riordan2011} rigorously proves that the explosive percolation
is always a continuous transition
unless a global dynamic is applied~\cite{Araujo2010,Chen2011}.
It is now a common belief that the continuity is an essential
feature of explosive percolation and of its variants~\cite{Boccaletti2016}.
However, the underlying mechanism for the anomalous phenomena,
particularly those misinterpreted as signs of
discontinuous transition, remains elusive, and
a robust estimate of critical exponents is still challenging.

In this Letter, we perform a systematic and extensive
simulation of the Achlioptas process in spatial dimensions
from $2$ to $6$ and on random graphs.
Given a finite system of sites $N$,
we define the pseudo-critical point in each random dynamic process
as the time step $\mathcal{T}_N$ when the incremental
size of the largest cluster reaches its maximum~\cite{Lee2011,Nagler2011,Fan2020},
and calculate its average
$T_N \equiv \langle \mathcal{T}_N \rangle$ and variance
$\sigma_{\mathcal T} \equiv \sqrt{\langle \mathcal{T}_N^2 \rangle-
\langle \mathcal{T}_N \rangle^2}$.
We observe two scaling windows:  the distance of $T_N$ to
the thermodynamic percolation threshold $T_c$
is $T_N-T_c \sim {\mathcal O}(N^{-1/\nu_1})$
and the variance is $\sigma_{\mathcal T} \sim {\mathcal O}(N^{-1/\nu_2})$,
with exponent $\nu_2 > \nu_1$.
This is dramatically different from the standard continuous
transition, for which both $T_N-T_c$ and $\sigma_{\mathcal T}$
are within a single scaling window ${\mathcal O}(N^{-1/\nu})$.
In addition to two correlation-length exponents $\nu_1$ and $\nu_2$,
other rich critical behaviors occur.
In the broad scaling window ${\mathcal O}(N^{-1/\nu_2})$,
the fractal dimensions are different from that
in the narrow one ${\mathcal O}(N^{-1/\nu_1})$ including ${\mathcal T}_N$,
and they are asymmetric at the two sides of the criticality.
Moreover, at the super-critical side, the largest cluster
and the others have different fractal dimensions.

\begin{table*}
\caption{Results of percolation threshold $T_c$ and
critical exponents with respect to system volume $N$.
The systems are hypercubic lattice for finite
dimension $d$ and on random graph for $d =\infty$.
The fractal dimensions $d_f^\pm$ and $d_{f2}^+$ are
for the broad scaling window
$\mathcal{T}_{c}^{\pm}(N,\nu_2)$.} \label{t1}
\begin{ruledtabular}
\begin{tabular}{clllllll|lcc}
& \multicolumn{7}{c|}{Explosive percolation}       &  \multicolumn{3}{c}{Bond percolation~\cite{Paul2001,Wang2013a,Xu2013,Mertens2018,Zhang2021}}  \\
$d$ & \multicolumn{1}{c}{$T_c$}       &  \multicolumn{1}{c}{$1/\nu_1$}      &  \multicolumn{1}{c}{$1/\nu_2$}     &  \multicolumn{1}{c}{$d_f$}   & \multicolumn{1}{c}{$d_f^-$}   & \multicolumn{1}{c}{$d_f^+$}     &  \multicolumn{1}{c|}{$d_{f2}^+$} &  \multicolumn{1}{c}{$T_c$} &  \multicolumn{1}{c}{$1/\nu$} &  \multicolumn{1}{c}{$d_f$}   \\
\hline
$2$ &  $1.053129(5)$  &  $0.512(1)$   &  $0.484(4)$   & $0.979(1)$  & $0.949(3)$   &   $0.980(3)$   &  $0.93(2)$    & $1$ & $3/8$ &  $91/96$ \\
$3$ &  $0.966301(1)$   &  $0.595(8)$   &  $0.500(1)$     & $0.924(5)$   & $0.789(8)$   &   $0.943(3)$    &  $0.80(2)$  & $0.746435$ & $0.380$  & $0.841$ \\
$4$ &  $0.936642(2)$   &  $0.67(4)$   &  $0.501(2)$     & $0.90(1)$   & $0.705(3)$   &   $0.930(7)$    &  $0.70(1)$ & $0.640525$ & $0.365$  & $0.761$ \\
$5$ &  $0.923283(1)$   &  $0.71(3)$   &  $0.501(1)$     & $0.915(5)$   & $0.68(3)$   &   $0.93(2)$    &  $0.69(2)$  & $0.590857$ & $0.349$  & $0.705$ \\
$6$ &  $0.915853(1)$   &  $0.72(2)$   &  $0.499(2)$     & $0.921(8)$   & $0.66(2)$   &   $0.951(3)$    &  $0.70(4)$  & $0.565210$ & $1/3$  & $2/3$ \\
$\infty$ &  $0.8884491(2)$   &  $0.740(2)$   &  $0.503(3)$     & $0.935(1)$   & $0.657(3)$   &   $0.956(2)$    &  $0.710(5)$  & $1/2$  & $1/3$  & $2/3$
\end{tabular}
\end{ruledtabular}
\end{table*}

In the conventional ensemble with a fixed density of
occupied bonds, correct scaling predictions are difficult
due to the sophisticated and anomalous phenomena.
In contrast, the scaling behaviors in the event-based ensemble
are rather clean and enable us to determine with
a high precision the percolation threshold $T_c$
and a variety of critical exponents
in dimensions from $2$ to $6$ and on random graphs (Table~\ref{t1}).
As a result, a systematic comparison with the standard bond
percolation becomes possible, and the explosive feature of
the transition is clearly illustrated.
Since multiple scalings may exist in other statistical
systems, the event-based method can find wide applications.

\begin{figure}
\centering
\includegraphics[width=1.0\columnwidth]{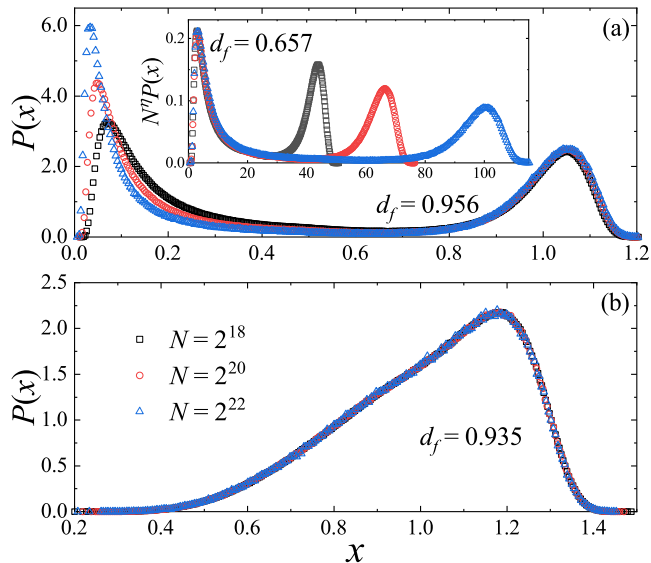}
\caption{Illustration of ensemble inequivalence by the
largest-cluster probability distribution $P(x={\mathcal C}_1/N^{d_f})$
on random graphs, where the value of $d_f$ is tuned to
achieve data collapse  for different volumes $N$.
(a) is for the conventional ensemble at $T_c$ and (b) is
for the event-based ensemble at $\mathcal{T}_N$.
The inset of (a) is for data collapse for the left hump,
where the height is rescaled by factor $N^\eta$ with $\eta=0.08$.
} \label{f1}
\end{figure}

\textit{Method and Observables}.--By adopting the basic
product rule and a flatten tree data structure,
at each time step $t$ we insert one of two randomly chosen bonds
giving the smaller product of cluster sizes,
and record the sizes of clusters as
${\mathcal C}_i \, (i=1,2,\cdots)$
and the one-step increasing gap of the largest cluster
$\Delta_{\mathcal C} (t) \equiv {\mathcal C}_1 (t)-{\mathcal C}_1 (t-1)$.
Roughly speaking, as $t$ evolves, the gap $\Delta_{\mathcal C}$
first increases in the sub-critical phase with small clusters,
peaks near the percolation threshold, and finally decreases
in the super-critical phase with a giant and dense cluster.
The pseudo-critical point ${\mathcal T}_N = t_1/N$ corresponds
to the time step $t_1$ of the event that
$\Delta_{\mathcal C}$ reaches its maximum.
At ${\mathcal T}_N$, various observables
are sampled, including the sizes of the largest and second
largest clusters (${\mathcal C}_1$ and ${\mathcal C}_2$),
and the probability distribution $P(x={\mathcal C}_1)$.
Their average values are calculated as $T_N, C_1$ and $C_2$.

The thermodynamic threshold $T_c$
is obtained by the least-squares fit of the $T_N$ data to
\begin{equation}
T_N=T_c+N^{-1/\nu_1} (a_0+a_1 N^{-\omega_1}+a_2N^{-\omega_2}+\cdots) \; ,
\label{eq:fit_Tc}
\end{equation}
where the terms with $\omega_i \, (i=1,2)$ account for finite-size
corrections. The results are shown in Table~\ref{t1}.
For convenience, the critical exponents are quoted with
respect to system volume $N$ rather than linear size $L$.

\textit{Ensemble Inequivalence}.--As an example of the anomalous
behaviors in the conventional ensemble,
we consider the probability distribution
$P(x={\mathcal C}_1)$ at $T_c=0.8884491$ for random graphs.
It is clear that $P(x)$ exhibits a bimodal
distribution (Fig.~\ref{f1}(a)).
Different fractal dimensions are used to obtain data collapse
for each hump, and the total probability for the left hump
asymptotically vanishes with a small exponent.
In contrast, Fig.~\ref{f1}(b) shows that,
at the pseudo-critical point ${\mathcal T}_N$,
$P(x)$ is unimodal and is well renormalized by
a single fractal dimension. This provides solid evidence for
the continuity of explosive percolation.

Moreover, neither of the two fractal dimensions
in Fig.~\ref{f1}(a) is identical to the correct value
$d_f \! =\! 0.935$ in Fig.~\ref{f1}(b).
Since the normalized distribution $P(x={\mathcal C}_1/N^{d_f})$
holds true in the $N \! \to \! \infty$ limit,
this inequivalance of two ensembles cannot vanish as $N$ increases,
and, thus, the $d_f$ value cannot be correctly
extracted from the conventional ensemble.

\begin{figure}
\centering
\includegraphics[width=1.0\columnwidth]{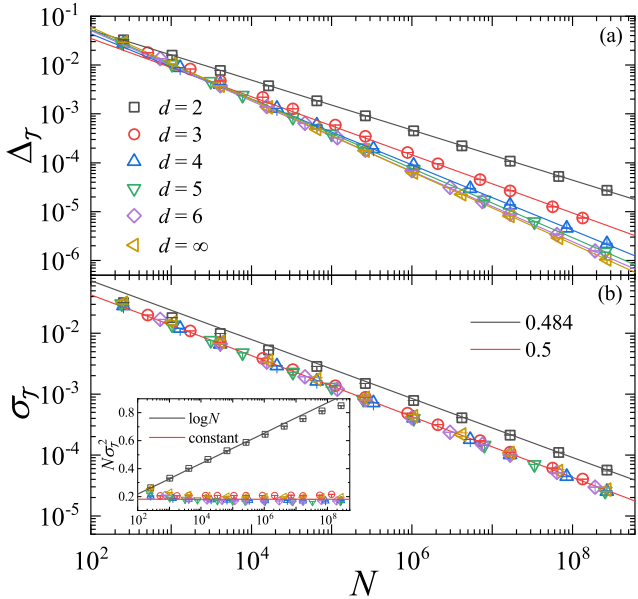}
\caption{Illustration of two scaling windows by
the pseudo-critical distance
$\Delta_{\mathcal T}\equiv\langle|\mathcal{T}_N-\mathcal{T}_{N,2}|\rangle$ (a)
and the variance $\sigma_{\mathcal T}$.
The exponent $1/\nu_2$ is consistent $1/2$ except for $d=2$,
while $1/\nu_1$ is $d$-dependent and larger than $1/2$.
The inset of (b) shows the linear-log plot of the specific-heat-like quantity
$N \sigma_{\mathcal T}^2$ as a function of system volume $N$.} \label{f2}
\end{figure}

\textit{Two scaling windows}.--
The observed ensemble inequivalence is counter-intuitive
and seems inconsistent with the standard finite-size
scaling (FSS) theory. Let us recall some basic ideas.
Near a critical point $T_c$, the diverging
correlation length, $\xi(T) \sim |T-T_c|^{-\nu_L}$,
is the only characteristic length scale
($\nu_L$ is the correlation-length exponent with respect to $L$),
and the scaling of a physical quantity ${\mathcal Q} (T)$
is then a function of $\xi$.
The FSS theory assumes that, for a finite system,
$\xi_L$ is saturated to be the order of linear size $L$
when approaching some pseudo-critical point $T_N$,
and accordingly, the FSS of ${\mathcal Q}$ reads
\begin{equation}
{\mathcal Q} (T,N) \sim N^y \, \tilde{\mathcal Q}
\left(N^{1/\nu} (T-T_c)\right) \; ,
\label{eq:FSS_Tc}
\end{equation}
where $\tilde{\mathcal Q}$ is a universal function,
and $\nu/d=\nu_L$ and $y$ are critical exponents.
The pseudo-critical point is $T_N \simeq  T_c + a N^{-1/\nu}$
from $\xi_L \! \sim \! L$, with $a$ some constant.
Thus, the FSS of ${\mathcal Q}$ near $T_N$
still follows Eq.~(\ref{eq:FSS_Tc}), with the
argument $(T-T_c) L^{1/\nu}$
in $\tilde{\mathcal Q}$ being shifted as $(T-T_N) N^{1/\nu}+a$.
This means that the observed ensemble equivalence should not
be attributed to the deviation of $T_N$ from $T_c$.

We now argue that the explosive percolation exhibits
two scaling windows: while the deviation
$T_N-T_c \sim N^{-1/\nu_1}$, the sample-to-sample fluctuation $\sigma_{\mathcal T}$
is of order ${\mathcal O} (N^{-1/\nu_2})$ with $\nu_2 > \nu_1$.
As a result, the fluctuation plays
an important role in the fixed-bond-density ensemble.

In actual simulations, a set of pseudo-critical points,
${\mathcal T}_{N,\ell}$ ($\ell =1,2,\cdots$),
is sampled from events that the gap
$\Delta_{\mathcal C}$ reaches its $\ell$-th maximal value
(${\mathcal T}_N$ is simply ${\mathcal T}_{N,1}$).
The asymptotic behavior $T_{N,\ell}-T_c \sim N^{-1/\nu_1}$
is confirmed for all $\ell$.
To better estimate $1/\nu_1$,
we fit the data of $\Delta_{\mathcal T} \equiv
\langle |{\mathcal T}_{N}-{\mathcal T}_{N,2}| \rangle$ to
\begin{equation}
\Delta_{\mathcal T} (N) = N^{-1/\nu_1} (a_0+a_1
N^{-\omega_1}+a_2 N^{-\omega_2}+\cdots) \; ,
\label{eq:fit_deviation}
\end{equation}
which significantly eliminates the sample-to-sample fluctuation.
Also, the variance $\sigma_{\mathcal T}$ of ${\mathcal T}_N$
is fitted to Eq.~(\ref{eq:fit_deviation})
with $\nu_1$ being replaced by $\nu_2$.
The results in Table~\ref{t1} clearly support
our argument of two scaling windows with $\nu_2 > \nu_1$,
which is further illustrated by Fig.~\ref{f2}.

Unlike the $d$-dependent value of $\nu_1$,
$\nu_2$ is consistent with $2$ for $d \geq 3$.
According to the central-limit theorem,
the distribution of an intensive quantity is of Gaussian type
for sufficiently large $N$ and its variance is of order ${\mathcal O}(N^{-1/2})$.
As shown in the inset of Fig.~\ref{f2}(b), for $d \geq 3$ the specific-heat-like
quantity $N \sigma_{\mathcal T}^2$ quickly converges to a constant as $N$ increases,
suggesting that the central-limit theorem might hold true.
In two dimensions, since $1/\nu_1 \! = \! 0.512(1)$
is close to $1/\nu_2 \! = \! 0.484(4)$,
we suspect that the small deviation $1/2-1/\nu_2
\approx 0.016$ is due to finite-size effects
which are not adequately accounted for in the fit.
Indeed, Fig.~\ref{f2}(b) indicates that, as $N$ increases,
the slope of $\sigma_{\mathcal T}$ in log-scale increases toward $1/2$.

\begin{figure}
\centering
\includegraphics[width=1.0\columnwidth]{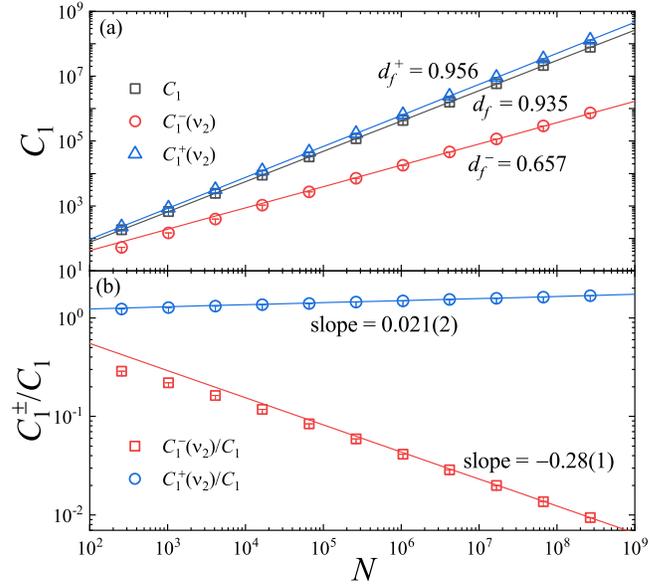}
\caption{Illustration of multiple fractal dimensions
on random graphs by the largest-cluster size $C_1$
at ${\mathcal T}_N$ and $C_1^{\pm}$ at
${\mathcal T}_N^{\pm } (\nu_2) = {\mathcal T}_N \pm N^{-1/\nu_2}$.
The ratios $C_1^{\pm}/C_1$ in (b) clearly show
$d_f^{\pm } \neq d_f$.} \label{f3}
\end{figure}

\textit{Multiple and fractal dimensions}.--
The $C_1$ and $C_2$ data at ${\mathcal T}_N$ are well described by
\begin{equation}
C =N^{d_f} (a_0+a_1
N^{-\omega_1}+a_2 N^{-\omega_2}+\cdots) \; ,
\label{eq:fit_C}
\end{equation}
and give a high-precision determination of the fractal
dimension $d_f$ for all dimensions (Table~\ref{t1}).
For each random Achlioptas process,
we also sample ${\mathcal C}_1^{\pm} (\nu_1)$ respectively
at $\mathcal{T}_N^\pm (\nu_1) \equiv \mathcal{T}_N
\pm a N^{-1/\nu_1}$ by simply taking $a=1$.
It is observed that, irrespective of dimension $d$,
the $C_1^{\pm} (\nu_1)$ data are also well described
by Eq.~(\ref{eq:fit_C}) and the $d_f$ values are
consistent with those in Table~\ref{t1}.
This means that, in the whole narrow scaling window
${\mathcal O} (N^{-1/\nu_1})$, the sample-to-sample fluctuation
is suppressed and the scaling behaviors are rather clean.
In contrast, it is difficult to obtain
a reliable estimate of $d_f$ in the conventional ensemble,
as implied by Fig.~\ref{f1}(a) for random graphs,
where the correct value $d_f=0.935(1)$ cannot be extracted.

We then explore geometric properties
within the broad scaling window ${\cal O} (N^{-1/\nu_2})$
by sampling at $\mathcal{T}_N^\pm (\nu_2)
\equiv \mathcal{T}_N  \pm N^{-1/\nu_2}$.
While Eq.~(\ref{eq:fit_C}) can still describe the data of
$C_1^{\pm} (\nu_2)$, the values of $d_f^{\pm}$ are clearly
different from those at ${\mathcal T}_N$ (Table~\ref{t1}).
This implies the emergence of multiple fractal dimensions.
On random graphs, the $C_1^{\pm} (\nu_2)$ data are shown in Fig.~\ref{f3}(a)
and their ratios to $C_1 ({\mathcal T}_N)$ are given in Fig.~\ref{f3}(b).
The value $d_f^-\!=\!0.657(3)$ at the sub-critical side
explains the appearance of the left hump in Fig.~\ref{f1}(a),
and $d_f^+ \!= \! 0.956(3)$ at the super-critical side
agrees well with the right hump.
In other words, in the conventional ensemble,
the bimodal feature of $P(x)$ is solely
from the broad scaling window ${\mathcal O}(N^{-1/\nu_2})$,
and the correct critical scaling at ${\mathcal T}_N$
is hidden in the sophisticated mixing effects.

\begin{figure}
\centering
\includegraphics[width=1.0\columnwidth]{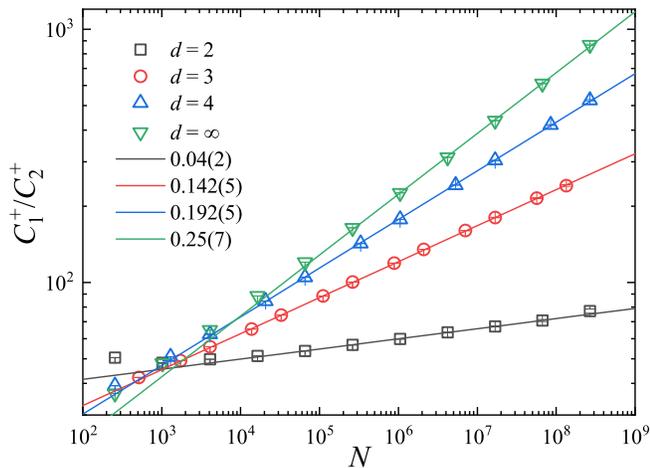}
\caption{Illustration of two length scales at
$\mathcal{T}_{N}^+ = \mathcal{T}_{N}+N^{-1/\nu_2}$
by size ratio $C_1^+/C_2^+$ for the largest and the second
largest cluster.}
\label{f4}
\end{figure}

\textit{Asymmetric fractal dimensions}.--
Another important feature in Fig.~\ref{f3}
is that, in the broad window $\mathcal{O}(N^{-1/\nu_2})$,
the FSS of the largest cluster $C_1$ is asymmetric
at the two sides of ${\mathcal T}_N$.
Namely, the fractal dimensions, $d_f^-$ and $d_f^+ $,
take different values, irrespective of $d$ (Table~\ref{t1}).

We also sample size ${\mathcal C}_2^{\pm}$
of the second largest cluster at ${\mathcal T}_N^{\pm} (\nu_2)$,
and fit the $C_2^{\pm}$ data by Eq.~(\ref{eq:fit_C}).
At sub-critical side, the fractal dimension $d_{f2}^-$
is consistent with $d_f^-$ for the largest cluster.
However, at super-critical side, $d_{f2}^+$ value is
significantly smaller than $d_f^+$ (Table~\ref{t1}).
This means that, at ${\mathcal T}_N^+ (\nu_2)$, the largest and
the second largest cluster have different length scales,
as illustrated by ratio $C_1^+/C_2^+$ in Fig.~\ref{f4}.

In fact, when the event-based sampling is taken at
$\mathcal{T}_N^\pm(\nu)$ with exponent $\nu_1<\nu<\nu_2$,
the fractal dimensions, $d_{f}^\pm$ and $d_{f2}^\pm$,
are found to continuously vary with exponent $\nu$
(see Supplemental Material).
This gives further sophistication of scaling behaviors
in the conventional ensemble.

%\textit{Cluster size distribution}.--

% We find that the explosive percolation always shows
% an ensemble-independent Fisher exponent $\tau=1+1/d_f$,
% where $d_f$ is the fractal dimension obtained by the event-based
% sampling at $\mathcal{T}_{c}(N)$.
% The multiple and asymmetric scaling can be only recognized
% from the power-law decay part of the cluster size distribution,
% see Supplemental Material.
% This further turns up evidence that the event-based ensemble
% succinctly captures the intrinsic nature of
% the explosive percolation, while the critical phenomena
% in the conventional ensemble are infiltrated with
% some incidental scaling features.

\textit{Discussion and Outlook}.--
By an event-based method and extensive simulations,
we find that the Achlioptas process for any $d$
has two scaling windows: the distance of the average
pseudo-critical point $T_N$ from the percolation
threshold $T_c$ and the standard deviation of
${\mathcal T}_N$ are characterized by two distinct exponents
$\nu_1$ and $\nu_2$.
Based on numerical evidences, we assume that, at least for $d \geq 3$,
the fluctuation of ${\mathcal T}_N$ is of Gaussian type and thus $\nu_2=2$.
Finite-size scaling (FSS) behaviors of
quantities, sampled at the random point ${\mathcal T}_N$ and
its narrow scaling window ${\mathcal O}(N^{-1/\nu_1})$,
are rather clean and can be described by
the standard FSS ansatz like Eqs.~(\ref{eq:fit_Tc})-(\ref{eq:fit_C}).
As a result, we determine with a high precision
the percolation threshold $T_c$ and various critical exponents
from $d=2$ to 6 and on random graphs (Table~\ref{t1}).

By a systematic comparison with the standard bond
percolation (BP) model,
we reveal the explosive but continuous nature of the explosive
percolation (EP), especially for $d \ge 3$.
In two dimensions, $T_c ({\rm EP}) \! \approx \! 1.053$
is slightly delayed about $5\%$ of $T_c ({\rm BP}) \! = \! 1$, and
the fractal dimensions $d_f({\rm EP})$ and $d_f({\rm BP})$
are close to each other.
Nevertheless, the explosive transition is clearly sharper
as seen from exponent $1/\nu$.
As $d$ increases, the gap $T_c ({\rm EP})-T_c ({\rm BP})$
becomes larger, and the explosive feature becomes
clearer as reflected by the increasing value of $1/\nu_1$.
Interestingly,  it seems that $1/\nu$ for
standard percolation reaches a maximum around $d=3$,
while $d_f$ for explosive percolation has a minimum around $d=4$.

Within the broad scaling window ${\mathcal O}(N^{-1/\nu_2})$,
rich phenomena are observed,
including multiple and asymmetric fractal dimensions
and two length scales at the super-critical side.
Note that multiple scaling windows are also observed in
the high-dimensional Ising model,
which are interpreted as the simultaneous existence
of multiple fixed points in the language of renormalization
group~\cite{Wittmann2014,Zhou2018,Fang2021,Fang2022}.
However, a theoretical understanding is
still needed for explosive percolation.
A possible scenario is that,
outside the narrow window ${\mathcal O}(N^{-1/\nu_1})$,
the rich behaviors are effectively
the crossover phenomena from finite systems to
thermodynamic limit. Even though,
quantitative predictions are desired for
the multiple fractal dimensions in Table~\ref{t1}.

Our work is of practical importance.
First of all, we point out that the anomalous phenomena,
previously observed in the conventional ensemble
of fixed bond density, are actually the mixing effects
of multiple fractal structures.
In particular, the bimodal probability distribution
is not a sign of first-order phase transition but
from the asymmetric scalings above and below ${\mathcal T}_N$.
Second, we point out that, in the conventional ensemble,
correct scaling predictions are difficult and
reliable estimates of critical exponents are challenging.
Finally, we point out that the event-based method
can find broad applications. An immediate and important application
is to address the controversial debate whether
the explosive percolation universality should depend
on the bond-inserting rules adopted in the Achlioptas process.
Large sample-to-sample fluctuations can widely exist in
systems like disordered ones~\cite{Pazmandi1997,Bernardet2000},
it is certain that the event-based method can serve as
a powerful tool in these cases.
Further, even for usual equilibrium statistical systems,
the event-based method has its own advantage
by sampling the pseudo-critical point ${\mathcal T}_N$
as some random event.
In contrast, in traditional simulations,
$T_N$ is frequently defined as the peak position of
susceptibility or specific heat, and
the location of $T_N$ needs a large number of samples
and suffers from the intrinsic mathematical difficulty
of finding a maximum point.

The research was supported by the Science and Technology
Committee of Shanghai (Grant No.~20DZ2210100), the National Key
R\&D Program of China (Grant No.~2018YFA0306501).

\bibliography{ref}

\appendix
\newpage
\setcounter{figure}{0}
\renewcommand*{\thefigure}{S\arabic{figure}}
\onecolumngrid

\begin{center}
{\large \bf Supplemental Material for ``Multiple and asymmetric scalings in explosive percolation"}
\end{center}

\subsection{Scalings at $\mathcal{T}_{N,\ell}$ for different $\ell$}

\begin{figure}
\centering
\includegraphics[width=1.0\columnwidth]{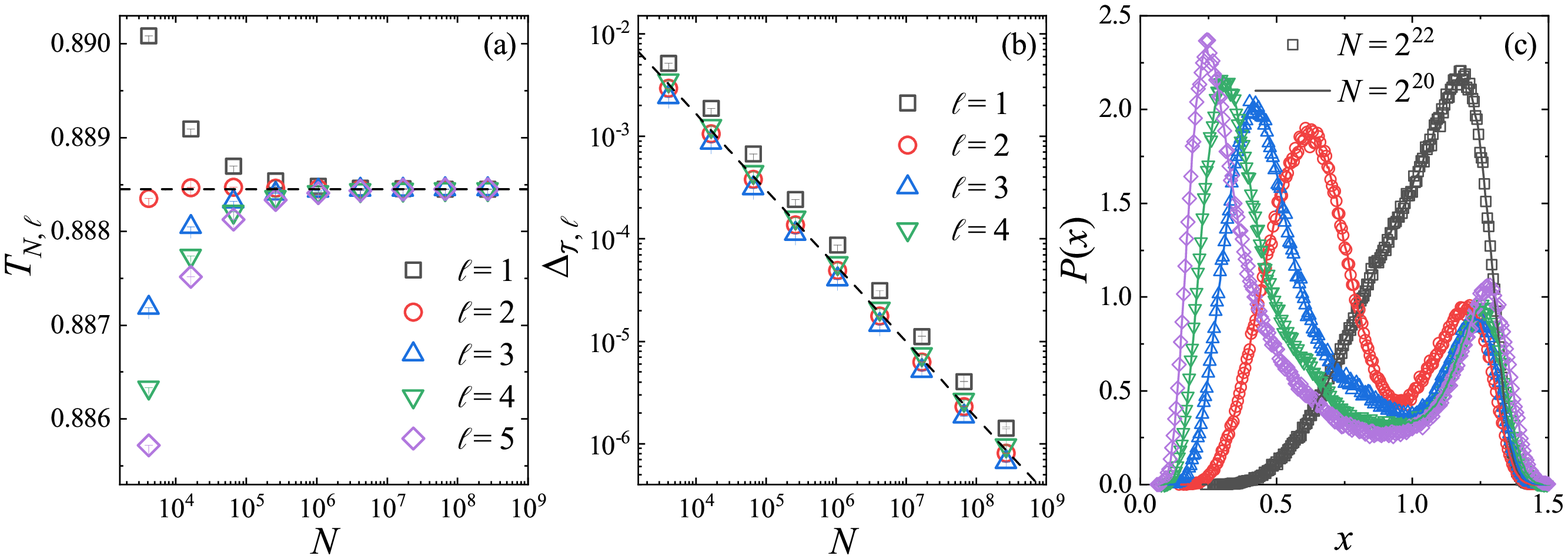}
\caption{(a) The pseudo-critical point $\mathcal{T}_{N,\ell}$ as a function of the system volume $N$ for different $\ell$. The black dashed line indicates the critical point $T_c=0.8884491$. (b) The distance between two pseudo-critical points $\Delta_{\mathcal{T},\ell} \equiv\langle |\mathcal{T}_{N,\ell} -\mathcal{T}_{N,\ell+1}|\rangle$ as a function of the system volume $N$ for different $\ell$. The dashed line shows a power law with a negative slope $-1/\nu_1=-0.74$. (c) The probability distribution of $\mathcal{C}_1$ at pseudo-critical point $\mathcal{T}_{N,\ell}$ for different $\ell$. Here $P(x)$ is the probability density function of $x=\mathcal{C}_1/N^{d_f}$ with $d_f=0.935$. The lines and scatters are for the systems of volumes $N=2^{20}$ and $N=2^{22}$, respectively. Here, the system is random graph.} \label{fs1}
\end{figure}

As stated in the main text, the pseudo-critical point $\mathcal{T}_{N,\ell}$ can be defined by the event that the one-step increasing gap of the largest cluster $\Delta_C$ reaches its $\ell$-th maximal value. All the averages $T_{N,\ell}\equiv\langle\mathcal{T}_{N,\ell}\rangle$ for different $\ell$ satisfy the same finite-size scaling
\begin{equation}
T_{N,\ell}=T_c+N^{-1/\nu_1} (a_0+a_1N^{-\omega_1}+a_2N^{-\omega_2}+\cdots).  \nonumber
\end{equation}
From Fig.~\ref{fs1} (a), we can find that the pseudo-critical point $T_{N,\ell}$ for different $\ell$ approaches to the critical point $T_c$ from different directions, i.e., $a_0$ can be either positive or negative. Figure~\ref{fs1} (b) shows that the distance between two pseudo-critical points $\Delta_{\mathcal{T},\ell} \equiv \langle |\mathcal{T}_{N,\ell} -\mathcal{T}_{N,\ell+1}|\rangle$ displays an $\ell$-independent scaling. This further provides an evidence that the asymptotic behaviors of all the pseudo-critical points $T_{N,\ell}$ give the same scaling exponent $1/\nu_1$, but different $a_0$ and correction terms.

In our experience, the statistical error of the observable at $\mathcal{T}_{N,\ell}$ increases with $\ell$. This is mainly because $\mathcal{T}_{N,\ell}$ for $\ell>1$ can be located either below or above the criticality, where the observable could be very different. A direct evidence is the bimodal distribution of $\mathcal{C}_1$ at $\mathcal{T}_{N,\ell}$ for $\ell>1$, see Fig.~\ref{fs1} (c). A brutal average on these remarkably different observables certainly introduces a large statistical error. Thus, it is not a good choice to use the Monte Carlo data at $\mathcal{T}_{N,\ell}$ with a large $\ell$.

In fact, the distribution of $\mathcal{C}_1$ at $\mathcal{T}_{N,1}$ is not a clean unimodal distribution yet. As shown in Fig.~\ref{fs1} (c), it is clear that the distribution is not symmetric on the two sides of the peak. This indicates that more than one case could be involved at $\mathcal{T}_{N,1}$. We can use the relationship between $\mathcal{T}_{N,1}$ and $\mathcal{T}_{N,2}$ to divide the samples into two groups, i.e., $\mathcal{T}_{N,1}>\mathcal{T}_{N,2}$ and $\mathcal{T}_{N,1}<\mathcal{T}_{N,2}$. Then, the symmetric unimodal distribution of $\mathcal{C}_1$ can be found in both the two groups.

Different from the bimodal distribution at the critical point $T_c$ (see Fig.~\ref{f1} in the main text), the bimodal distribution of Fig.~\ref{fs1} (c) only shows a unique fractal dimension $d_f=0.935$. This is because all these pseudo-critical points $\mathcal{T}_{N,\ell}$ are located inside the narrow scaling window $\mathcal{O}(N^{-1/\nu_1})$.

\subsection{The reordered pseudo-critical point $\mathcal{T}_{N,\ell}'$}

To eliminate the bimodal distribution of observables at the pseudo-critical point $\mathcal{T}_{N,\ell}$, one can sort the pseudo-critical points $\mathcal{T}_{N,\ell}$ obtained in a single realization by their values before the statistics of observables. Specifically, in a single realization we can relabel the largest pseudo-critical point as $\mathcal{T}_{N,1}'$, the second largest one as $\mathcal{T}_{N,2}'$, and so on. Then, the statistics of observables can be done at these reordered pseudo-critical points $\mathcal{T}_{N,\ell}'$ over different realizations. Note that the result also depends on the total number of the pseudo-critical points used in the reorder process.

\begin{figure}
\centering
\includegraphics[width=1.0\columnwidth]{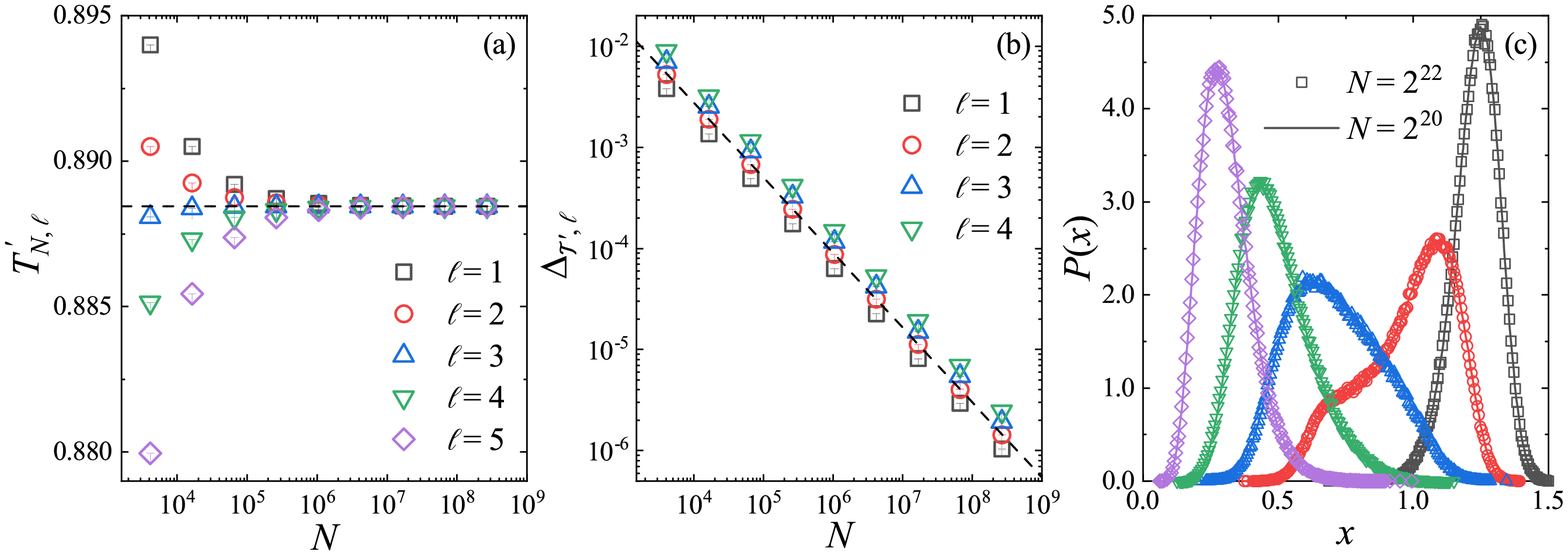}
\caption{(a) The reordered pseudo-critical point $\mathcal{T}_{N,\ell}'$ as a function of the system volume $N$ for different $\ell$. The black dashed line indicates the critical point $T_c=0.8884491$. (b) The distance of two reordered pseudo-critical points $\Delta_{\mathcal{T}',\ell} \equiv\langle |\mathcal{T}_{N,\ell}' -\mathcal{T}_{N,\ell+1}'|\rangle$ as a function of the system volume $N$ for different $\ell$. The dashed line shows a power law with a negative slope $-1/\nu_1=-0.74$. (c) The probability distribution of $\mathcal{C}_1$ at reordered pseudo-critical point $\mathcal{T}_{N,\ell}'$ for different $\ell$. Here $P(x)$ is the probability density function of $x=\mathcal{C}_1/N^{d_f}$ with $d_f=0.935$. The lines and scatters are for the systems of volumes $N=2^{20}$ and $N=2^{22}$, respectively. Here, the system is random graph.} \label{fs2}
\end{figure}

In Fig.~\ref{fs2}, we can find that the distributions of $\mathcal{C}_1$ for large and small $\ell$ become unimodal as expected. For a modest $\ell$, this reorder process cannot guarantee that the pseudo-critical point $\mathcal{T}_{N,\ell}'$ is always larger or smaller than the critical point, which results in a bimodal distribution. For a better result, one can sort more pseudo-critical points in the reorder process. Since this reorder process just performs a linear recombination of the observables found at $\mathcal{T}_{N,\ell}$, thus it does not change the scaling behavior, see Fig.~\ref{fs2}. In our experience, we cannot achieve a significantly improved fit result at the reordered pseudo-critical point $\mathcal{T}_{N,\ell}'$ for most cases. However, the reordered pseudo-critical point also has its advantages in dealing with some special cases, such as distinguishing the pseudo-critical points below and above the criticality.

\subsection{The asymptotic behavior of $T_N$ in different dimensions}

\begin{figure}
\centering
\includegraphics[width=1.0\columnwidth]{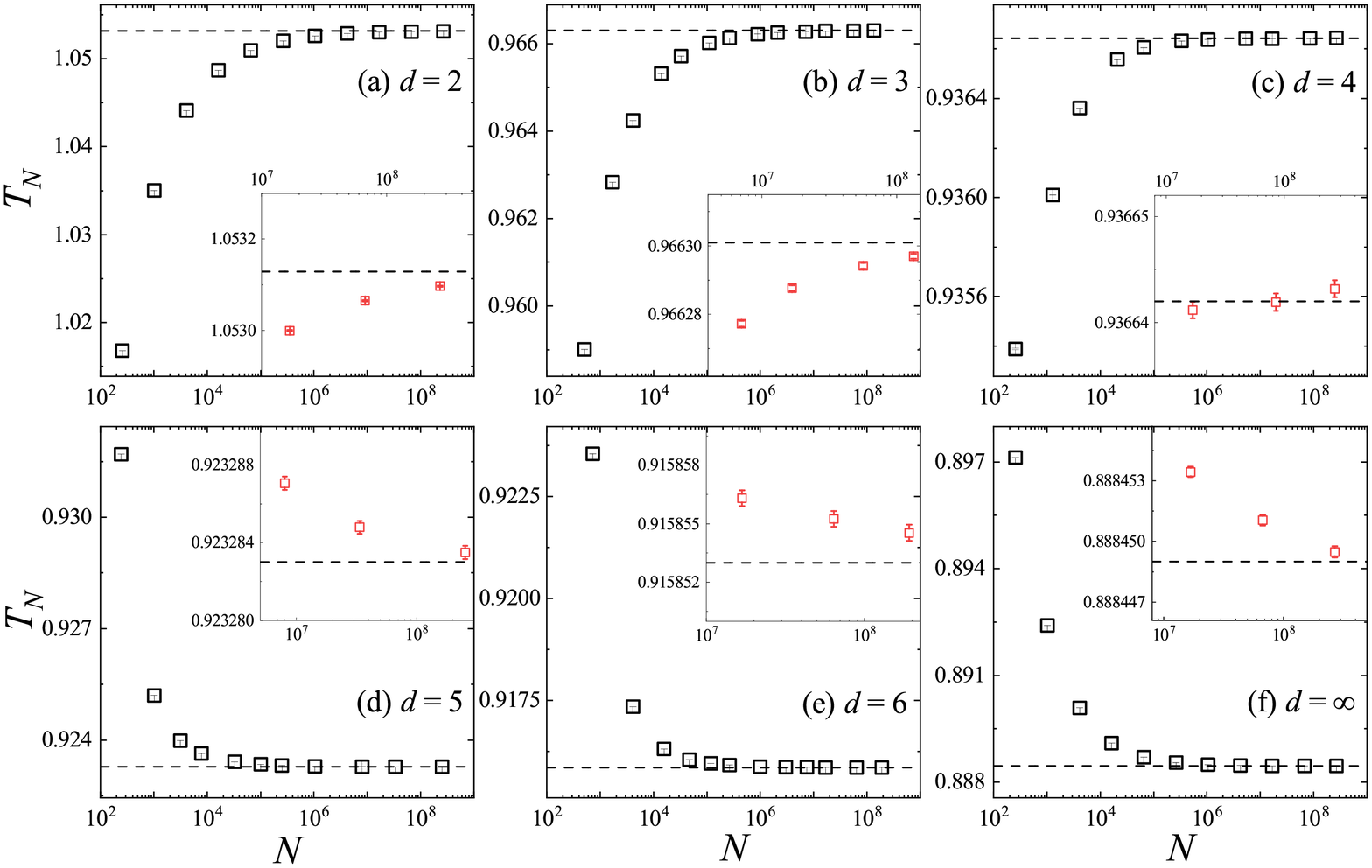}
\caption{The asymptotic behavior of the pseudo-critical point $T_N$ in different dimensions. The insets show the zooming figures for large systems. The dashed lines show the corresponding critical point $T_c$.} \label{fs3}
\end{figure}

In Fig.~\ref{fs3}, we show the asymptotic behavior of $T_N\equiv\langle\mathcal{T}_{N}\rangle$ for different dimensions. We can see that for $d<4$, the pseudo-critical point $T_N$ approaches to $T_c$ from below, while for $d>4$, the pseudo-critical point $T_N$ approaches to $T_c$ from above. For $d=4$, when the system is large enough, $T_N$ will be also larger than $T_c$.

\begin{figure}
\centering
\includegraphics[width=1.0\columnwidth]{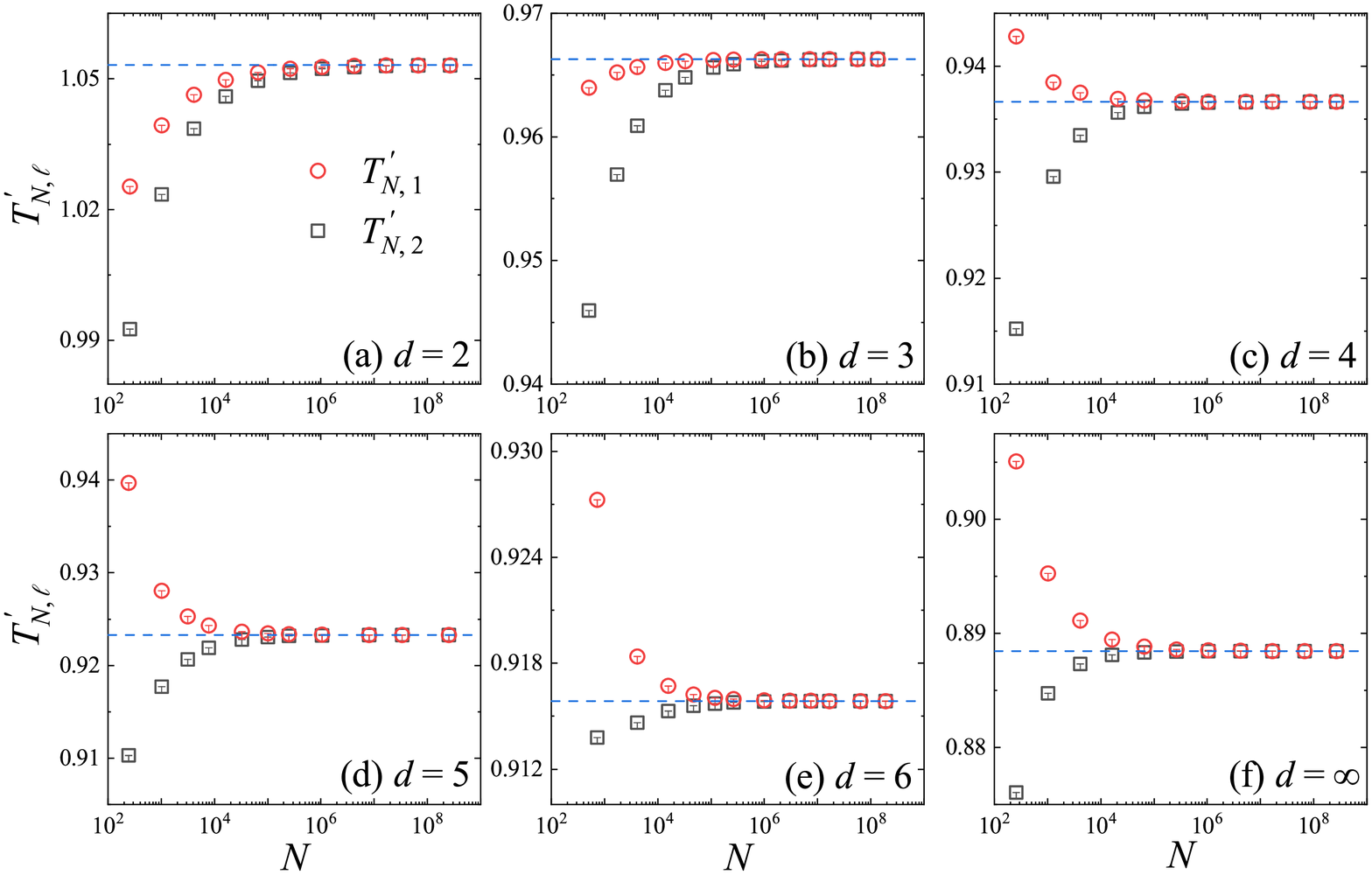}
\caption{The asymptotic behaviors of the reordered pseudo-critical points $T_{N,1}'$ and $T_{N,2}'$ in different dimensions. The dashed lines show the corresponding critical point $T_c$.} \label{fs4}
\end{figure}

For the reordered pseudo-critical point $T_{N,\ell}'$, similar phenomenon can also be observed, see Fig.~\ref{fs4}. We can find that $T_{N,2}'$ is always smaller than $T_c$, while $T_{N,1}'$ will be larger than $T_c$ for $d\geq4$. It should be noted that if we sort more pseudo-critical points, it is possible to find a $T_{N,2}'$ larger than $T_c$ even for $d<4$.

\subsection{Fractal dimensions $d_f^\pm(\nu)$ for different $\nu$}

To explore geometric properties within different scaling windows, we sample observables at $\mathcal{T}_N^\pm(\nu) \equiv\mathcal{T}_N\pm N^{1/\nu}$. Note that this is also an event-based sampling, since $\mathcal{T}_N^\pm(\nu)$ is also dependent on the occurrence of a special event. In order to realize this sampling, we first identify the pseudo-critical point $\mathcal{T}_N$ in a realization, and record the order in which bonds are occupied. Then, with the recorded bond sequence, one can reconstruct the percolation configurations at $\mathcal{T}_N^\pm(\nu)$. In this way, any observables at $\mathcal{T}_N^\pm(\nu)$ can be also obtained.

\subsubsection{$\nu=\nu_2$}

\begin{figure}
\centering
\includegraphics[width=1.0\columnwidth]{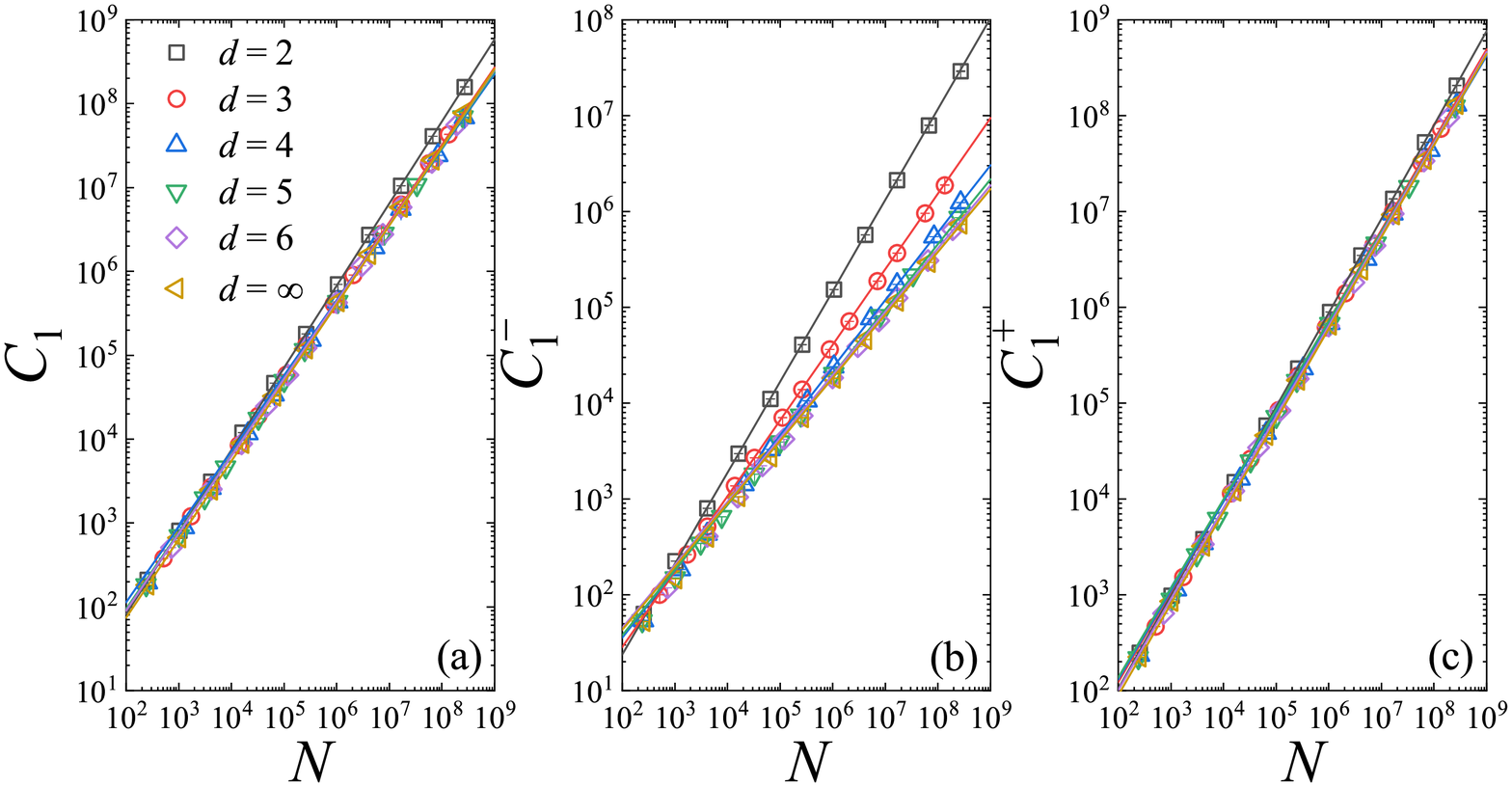}
\caption{The sizes of the largest clusters $C_{1}$ and $C_{1}^\pm(\nu_2)$ obtained by the event-based sampling as a function of the system volume $N$. The lines correspond to the fractal dimensions $d_f$ and $d_f^\pm(\nu_2)$ shown in Table \ref{t1} of the main text.} \label{fs5}
\end{figure}

\begin{figure}
\centering
\includegraphics[width=1.0\columnwidth]{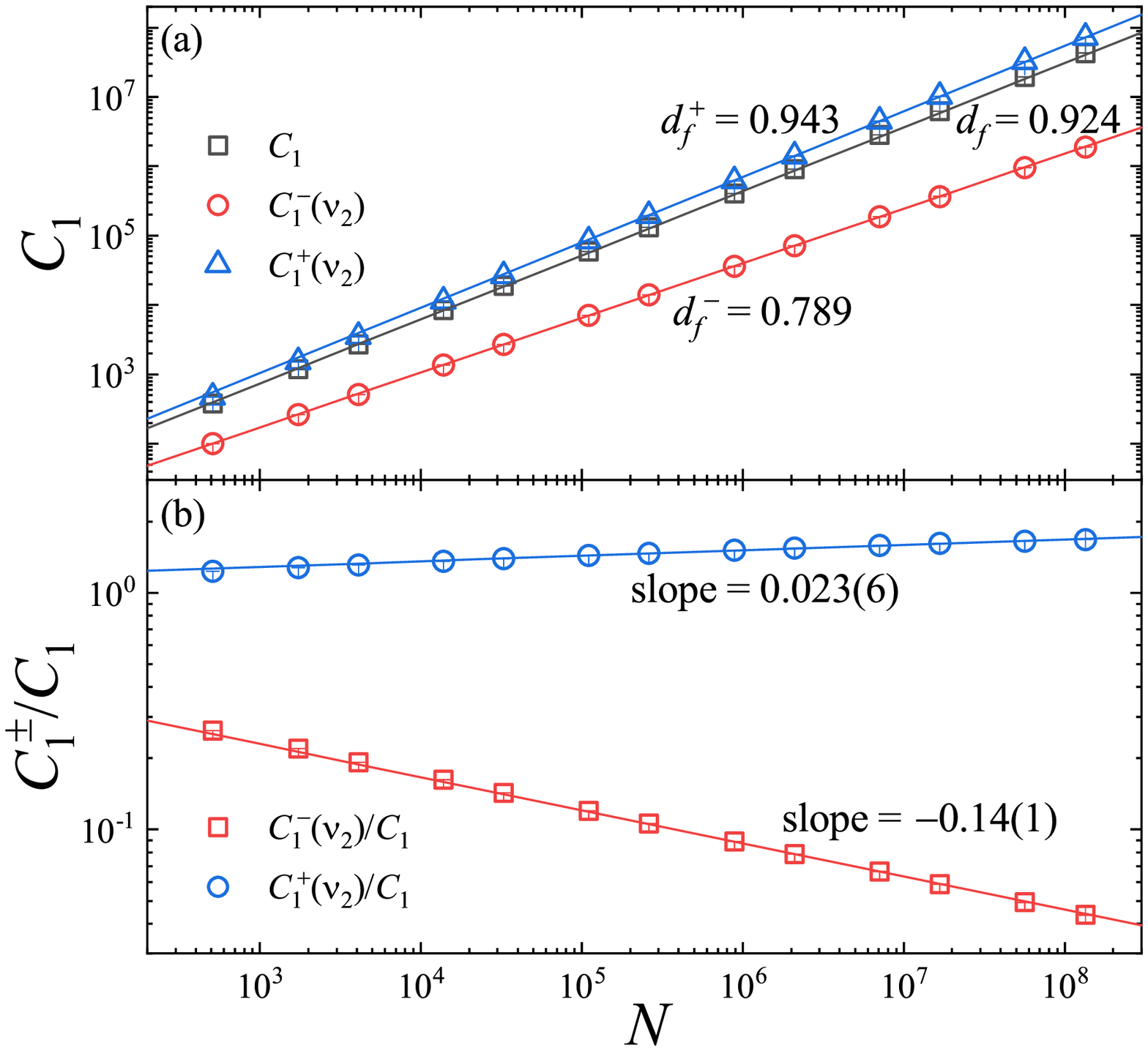}
\caption{Illustration of multiple fractal dimensions in dimension $3$ by the largest-cluster size $C_1$ at ${\mathcal T}_N$ and $C_1^{\pm}$ at ${\mathcal T}_N^{\pm } (\nu_2)$. The ratios $C_1^{\pm}/C_1$ in (b) clearly show $d_f^{\pm } \neq d_f$.} \label{fs6}
\end{figure}

In Fig.~\ref{fs5}, we show $C_{1}$ and $C_{1}^\pm(\nu_2)$ obtained by the event-based sampling as a function of the system volume $N$. We can find that $C_{1}$ and $C_{1}^\pm(\nu_2)$ show different fractal dimensions, indicating that there exists different fractal structures in the two scaling windows. In addition to Fig.~\ref{f3} in the main text, we also show the multiple fractal dimensions of $3$-dimensional systems in Fig.~\ref{fs6}.

\subsubsection{$\nu_1<\nu<\nu_2$ }

\begin{figure}
\centering
\includegraphics[width=1.0\columnwidth]{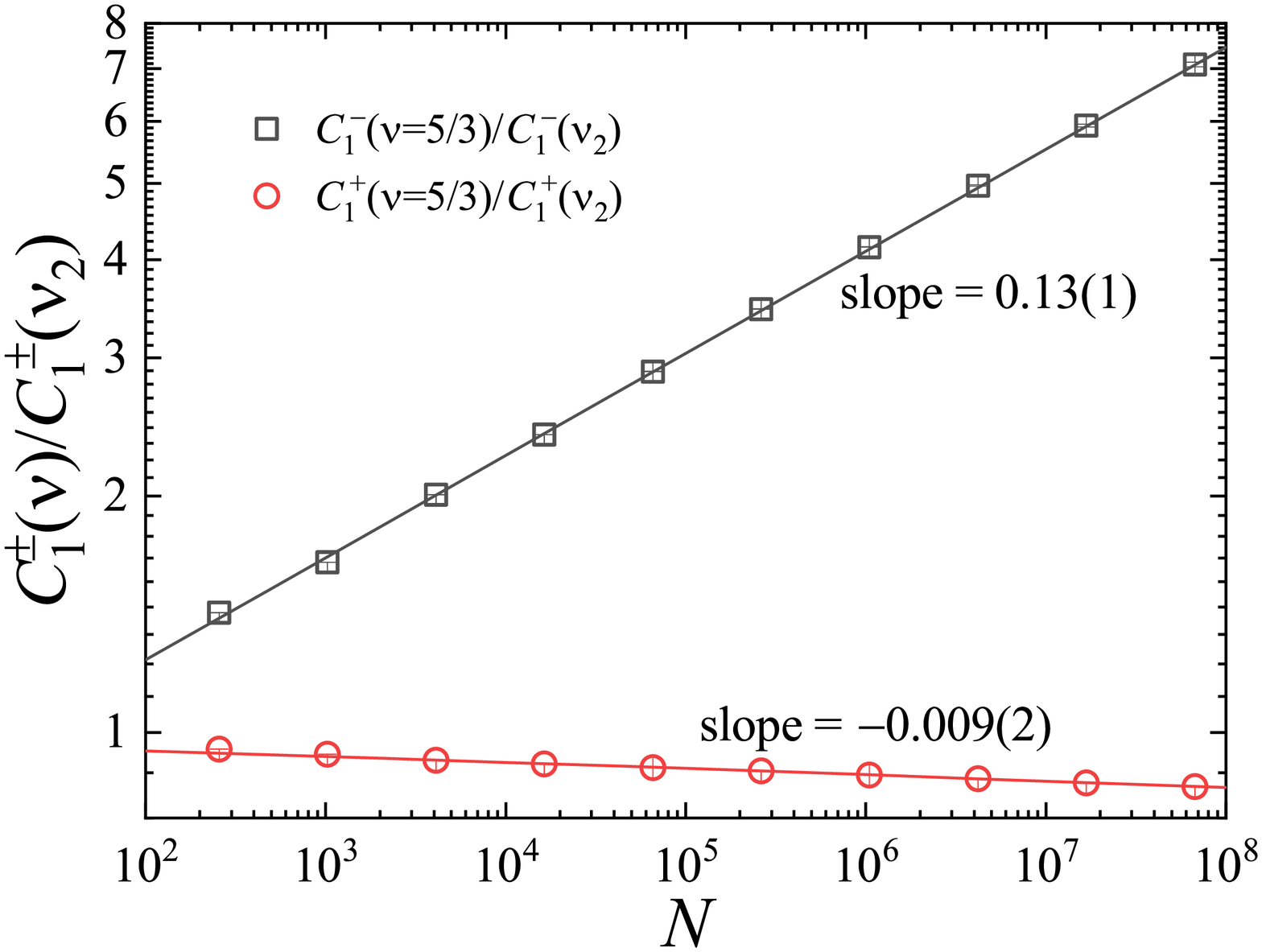}
\caption{The ratios $C_{1}^\pm(\nu=5/3)/C_{1}^\pm(\nu_2)$ as a function of the system volume $N$ on random graphs.} \label{fs7}
\end{figure}

In Fig.~\ref{fs7}, the ratios $C_{1}^\pm(\nu=5/3)/C_{1}^\pm(\nu_2)$ are plotted as a function of the system volume $N$. We can find that both the two ratios scale in a power law with the increase of the system volume. This means that at $\mathcal{T}_N^\pm(\nu=5/3)$ the largest cluster has a fractal dimension different from that at $\mathcal{T}_N^\pm(\nu_2)$. Similar phenomenon can also be found for other $\nu$. In that, we conjecture that outside the narrow scaling window $\mathcal{O}(N^{-1/\nu_1})$, there exists continuously variable fractal dimensions $d_f^\pm(\nu)$.

\subsection{Cluster size distribution}

\begin{figure}
\centering
\includegraphics[width=1.0\columnwidth]{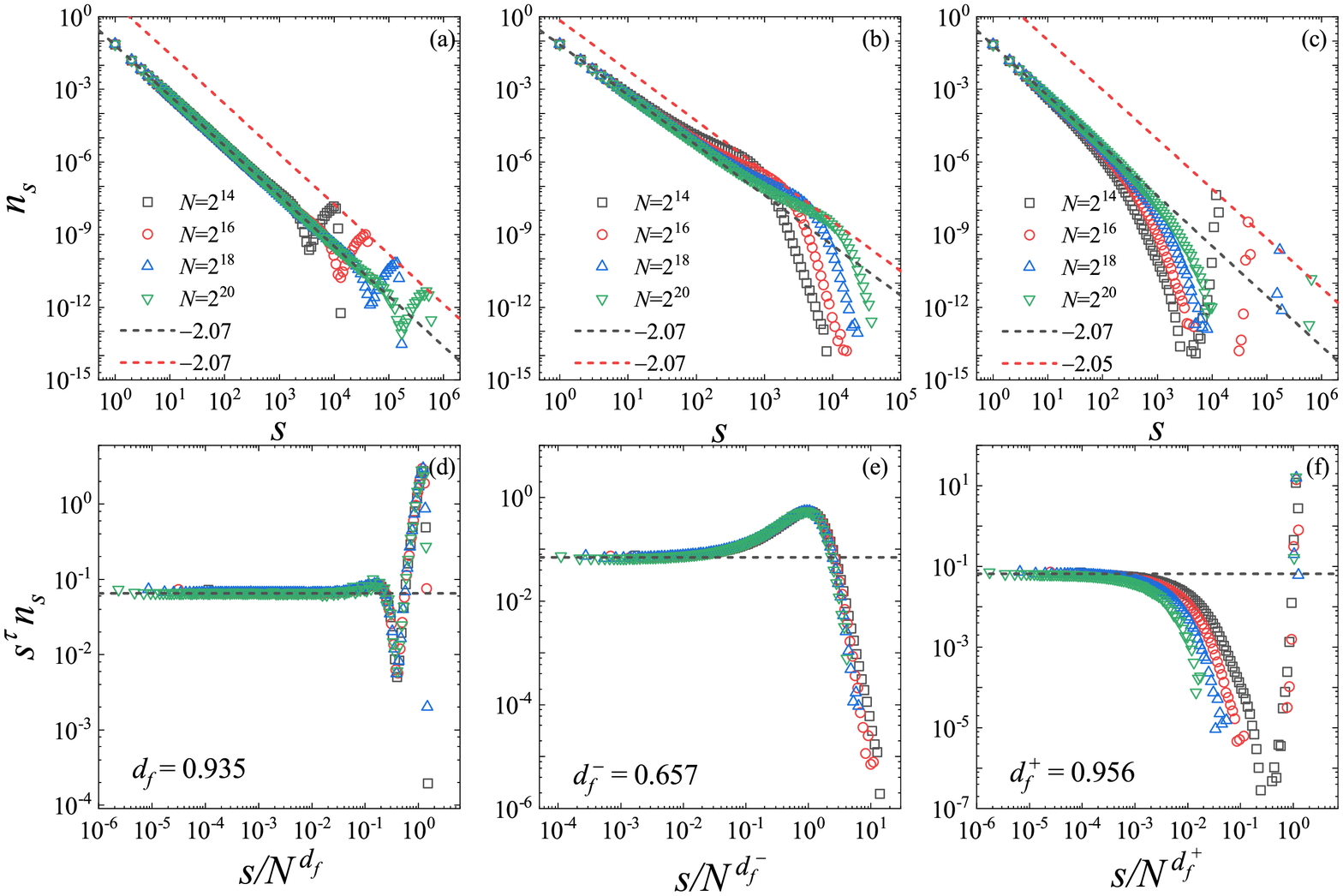}
\caption{The cluster size distributions $n_s$ on random graphs at (a) $\mathcal{T}_{N}$, (b) $\mathcal{T}_{N}^-(\nu_2)$, and (c) $\mathcal{T}_{N}^+(\nu_2)$. With the hyperscaling relation $\tau=1+1/d_f$, we know that $d_f=0.935$ and $d_f^{+}=0.956$ corresponds to $\tau=2.07$ and $2.05$, respectively. Sub-figures (d)-(f) are the corresponding rescaled forms of sub-figures (a)-(c) with $\tau=2.07$, respectively.} \label{fs8}
\end{figure}

\begin{figure}
\centering
\includegraphics[width=1.0\columnwidth]{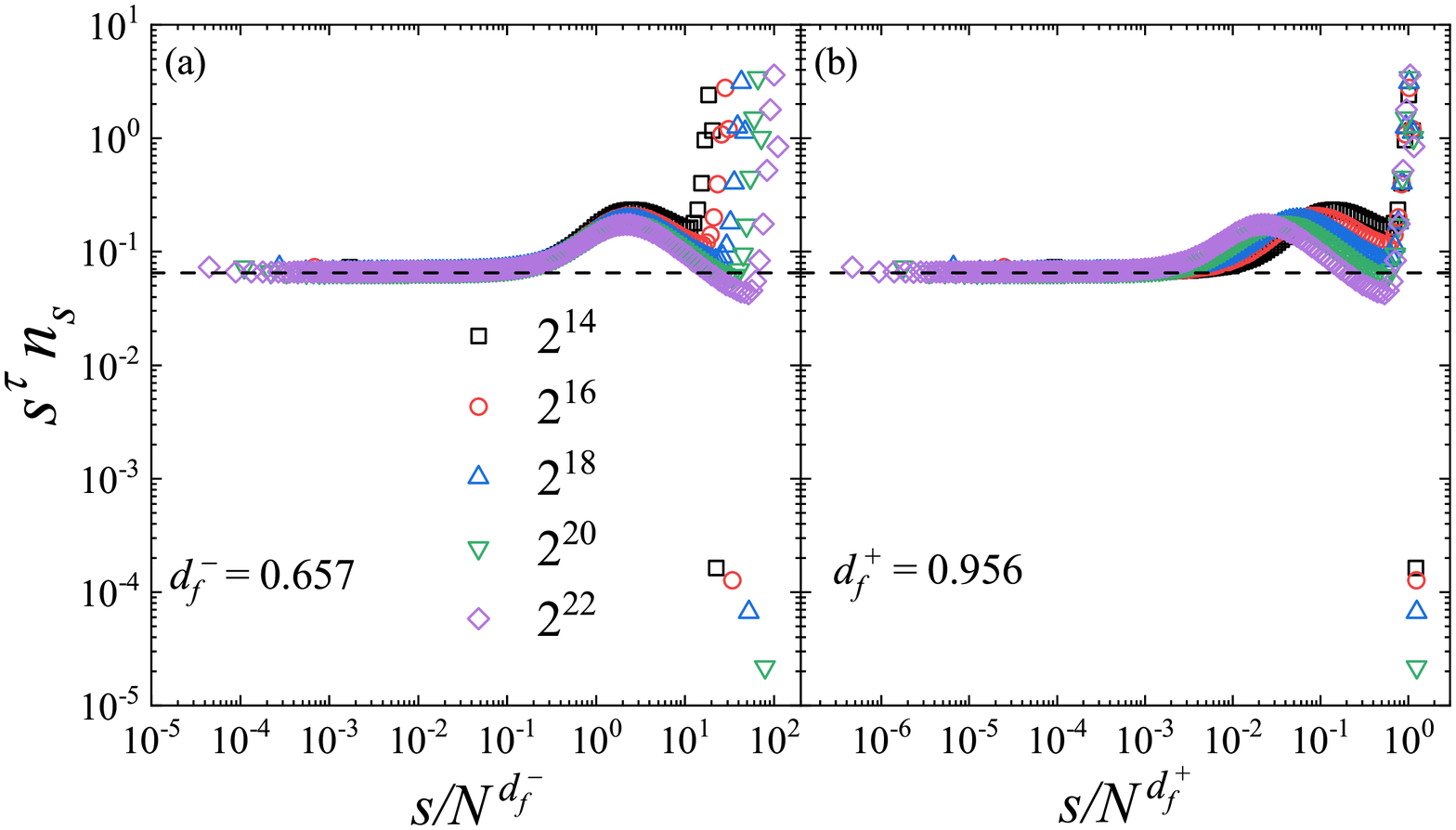}
\caption{The critical cluster size distribution $n_s$ on random graphs at $T_c=0.8884491$. Here, the Fisher exponent $\tau=1+1/d_f=1+1/0.935=2.07$ is used to rescale the cluster size distribution $n_s$.} \label{fs9}
\end{figure}

In Fig.~\ref{fs8}, we show the cluster size distributions $n_s$ at $\mathcal{T}_{c}$ and $\mathcal{T}_{c}^\pm(\nu_2)$ on random graphs. We can find that all the three cases show the same Fisher exponent $\tau=2.07$, which can be obtained by the hyperscaling relation $\tau=1+1/d_f$ with $d_f=0.935$. Moreover, different $d_f$ are required to show a good data collapse for the power-law decay part, see Figs.~\ref{fs8} (d)-(f). Even so, at $\mathcal{T}_{c}^\pm(\nu_2)$ (Figs.~\ref{fs8} (e) and (f)) we cannot find a single fractal dimension that can show a good data collapse over the entire range of the distribution $n_s$ as that at $\mathcal{T}_{c}$ (Fig.~\ref{fs8} (d)). This is mainly due to the multiple fractal dimensions outside the narrow scaling window $\mathcal{O}(N^{-1/\nu_1})$. Similar phenomenon can also be observed at $T_c$ shown in Fig.~\ref{fs9}.

\end{document}